\documentclass{iau}

\usepackage{amsmath}
\usepackage{graphicx}
\usepackage{multirow}

\usepackage{url}
\usepackage{adjustbox}
\usepackage{units}
\newcommand{\angstrom}{\mbox{\normalfont\AA}}

\begin{document}

\lefttitle{M. Brauner \textit{et al.}}
\righttitle{Chemical abundance inventory in phosphorus-rich stars}

\jnlPage{1}{7}
\jnlDoiYr{2025}
\doival{10.1017/xxxxx}
\volno{395}
\pubYr{2025}
\journaltitle{Stellar populations in the Milky Way and beyond}

\aopheadtitle{Proceedings of the IAU Symposium}
\editors{J. Mel\'endez,  C. Chiappini, R. Schiavon \& M. Trevisan, eds.}

\title{Chemical abundance inventory \\ in phosphorus-rich stars}



\author{Maren Brauner$^{1,2}$, Thomas Masseron$^{1,2}$, Marco Pignatari$^{3,4,5,6}$, D. Aníbal García-Hernández$^{1,2}$}

\affiliation{$^1$Instituto de Astrofísica de Canarias, C/Via Láctea s/n, E-38205 La Laguna,      Tenerife, Spain \\ email: \email{maren.brauner@iac.es} \\
$^2$Departamento de Astrofísica, Universidad de La Laguna, E-38206 La Laguna, Tenerife, Spain \\
$^3$Konkoly Observatory, Research Centre for Astronomy and Earth Sciences, HUN-REN, H-1121 Budapest, Konkoly Thege M. út 15-17, Hungary \\
$^4$CSFK, MTA Centre of Excellence, Budapest, Konkoly Thege Miklós út 15-17, H-1121, Hungary \\
$^5$Joint Institute for Nuclear Astrophysics - Center for the Evolution of the Elements \\
$^6$The NuGrid Collaboration, \url{http://www.nugridstars.org}}

\begin{abstract}
We provide an overview of the latest advances in the study of phosphorus-rich stars, covering their detailed chemical abundance analyses and innovative mining approaches. Following the discovery of 16 low-mass and low-metallicity stars rich in P, we expanded this sample by demonstrating that a recently identified group of Si-rich giants is also P-rich. A detailed abundance analysis was conducted on the near-infrared spectra from APOGEE-2 DR17, encompassing 13 elements. Subsequently, a similar analysis was performed on the optical UVES spectra of four P-rich stars, resulting in the abundance determination of 48 light and heavy elements. This comprehensive analysis further refined the chemical fingerprint of these peculiar stars, which was employed to evaluate the plausibility of various nucleosynthetic formation scenarios. In order to obtain a statistically more reliable chemical fingerprint in the future, we explored the use of unsupervised machine learning algorithms to identify additional P-rich stars in extensive spectroscopic surveys, such as APOGEE-2. The primary objective of this research is to identify the progenitor of these stars and determine whether current nucleosynthetic models require revision or if a completely new source of P in the Galaxy is responsible for the existence of the P-rich stars.

\end{abstract}


\maketitle

\section{Introduction}

Phosphorus, a key element in the chemical evolution of galaxies and biological systems, has its Galactic origins primarily attributed to production in massive stars. Current explanations of the Galactic origin of P focus on its production site in these stars. Models predicting the amounts of P in the stars of our Galaxy, as shown, for instance, in Fig. 1 of \cite{masseron20a} and Fig. 19 of \cite{kobayashi:20}, align reasonably well with observations at solar metallicities. However, significant discrepancies persist at lower metallicities, particularly around [Fe/H] = $\unit[-1]{dex}$. Coincidentally, this metallicity aligns with that of 16 phosphorus-rich stars discovered by \cite{masseron20a,masseron20b}. The origin of these P-rich stars remains an unresolved challenge in stellar nucleosynthesis. Their P content is between 10 and 100 times higher than that of the Sun relative to iron, while their masses are approximately $\unit[1]{M_\odot}$. Given their low mass, these stars could not have synthesized their high P content themselves. To address this discrepancy, we investigate the potential progenitors responsible for the anomalous P enrichment observed in these stars by: (1) extending the existing sample of P-rich stars using the same near-infrared survey in which these stars were initially discovered, namely APOGEE-2; (2) increasing the number of determined elemental abundances through the analysis of optical spectra; and (3) applying machine learning techniques to identify additional P-rich stars in large surveys, aiming for a statistically more reliable definition of their chemical fingerprint.



\section{Expanding the P-rich stars sample using APOGEE-2}\label{APOGEE}

\begin{figure}[t]
	\centering
	\includegraphics[width=0.4\textwidth]{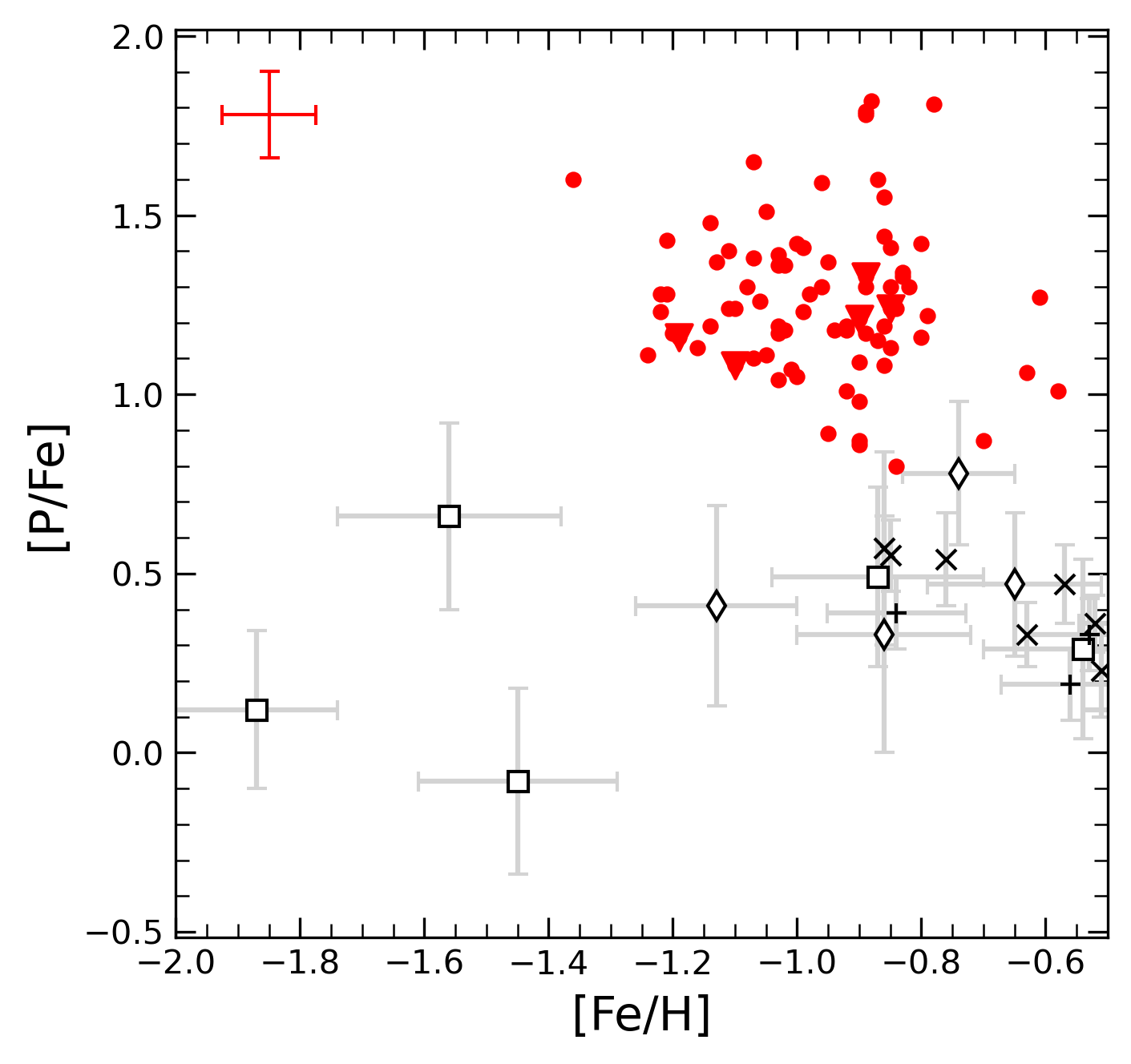} 
	\includegraphics[width=0.4\textwidth]{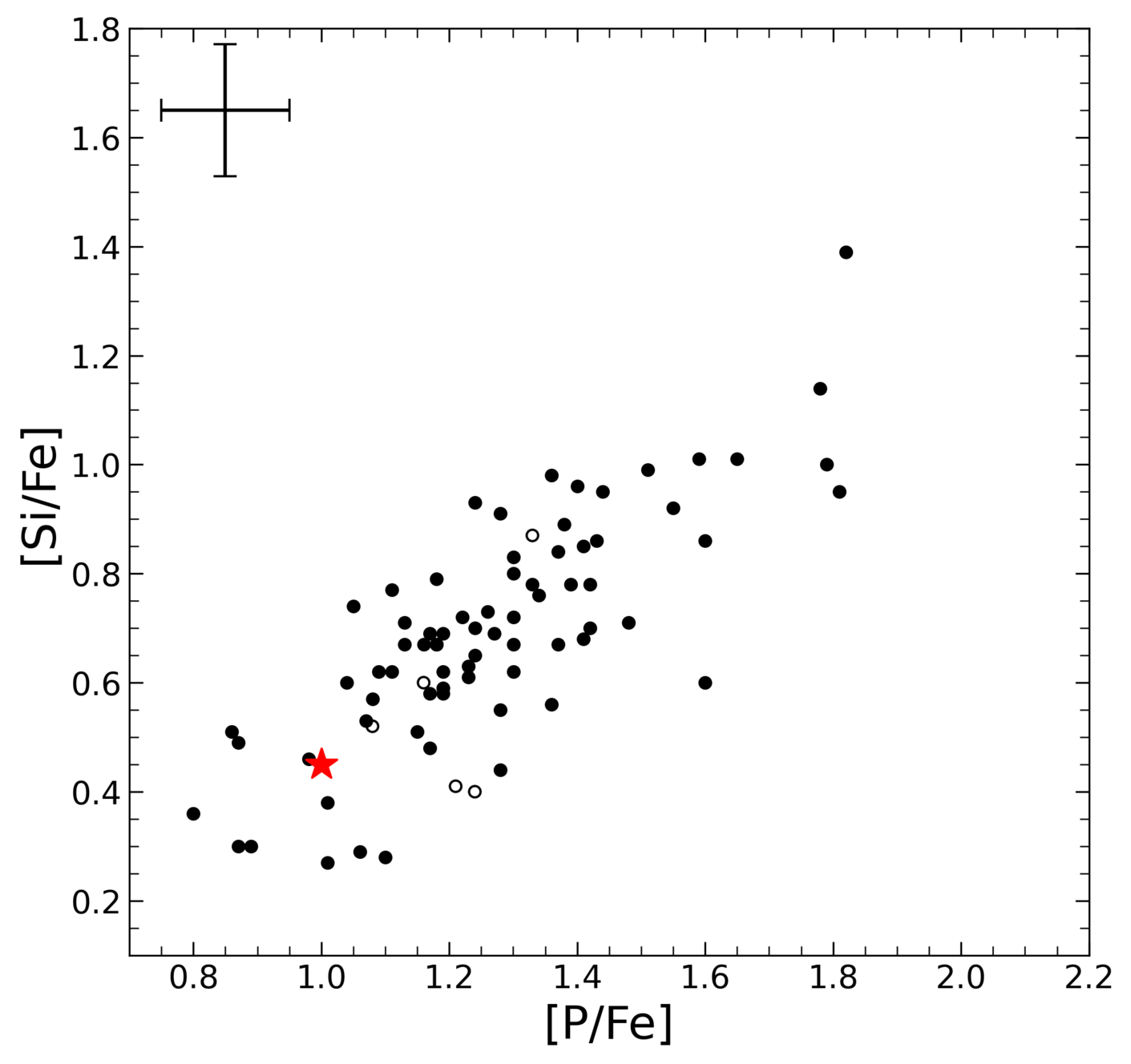} 
	\caption{Left panel: Abundance ratio [P/Fe] vs. metallicity [Fe/H] for the P-rich stars (red circles for measurements and red downward triangles for upper limits), compared to literature values (open black squares from \cite{jacobson14}, crosses from \cite{maas17,maas19}, diamonds from \cite{masseron20a}, and plus signs from \cite{caffau11,caffau19}). Right panel: [Si/Fe] vs. [P/Fe] for the P-rich stars. Solid circles correspond to real measurements for both Si and in P, while empty circles denote upper limits in P. The red star symbol highlights the P-rich star that is part of the globular cluster M4. Typical errors are indicated in the top-left corner of both panels.}
	\label{P_Fe_P_Si}
\end{figure}

\begin{figure}
	\centering
	\includegraphics[width=0.8\textwidth]{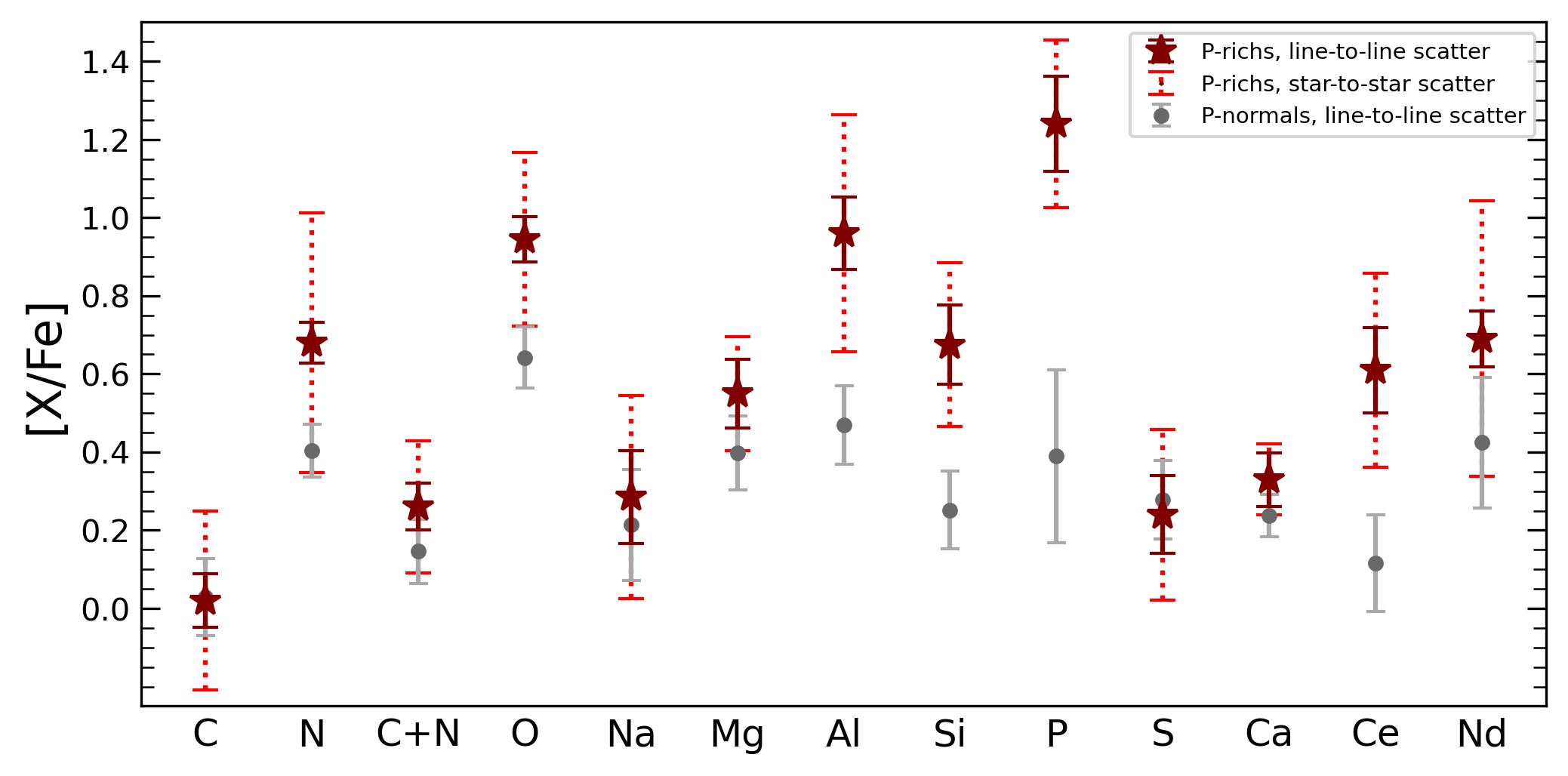} 
	\caption{Chemical abundance pattern of the P-rich star sample obtained using APOGEE-2 spectra. Dark red stars represent the median chemical abundance of the P-rich sample, while gray dots show the median chemical abundance of the background sample. Stars with upper limits were excluded from the median. Red and gray error bars denote the corresponding average line-to-line abundance scatter (typical standard deviation), which reflects the quality of the measurements. Dotted error bars represent the star-to-star abundance scatter of the sample, reflecting the spread of abundances. For P, the gray dot indicates the median abundance of the literature values shown in the left panel of Fig. \ref{P_Fe_P_Si}, while the gray error bar shows the mean error of the literature values.}
	\label{abu_summary}
\end{figure}

Apart from high amounts of P, \cite{masseron20a} showed that the P-rich stars are also enhanced in Si relative to iron. Therefore, a sample of Si-rich giants reported by \cite{fernandeztrincado20} served as a promising starting point for enlarging the sample of P-rich stars by determining the P abundance in these Si-rich stars. We performed a detailed 1D local thermodynamic equilibrium (LTE) abundance analysis on the high-resolution near-infrared (H-band) spectra from the latest data release (DR17) of the APOGEE-2 survey \cite[][]{abdurrouf22APOGEEDR17}. The wavelength range covered by APOGEE-2 includes two P I lines, located at $\unit[15711.522]{\angstrom}$ and $\unit[16482.932]{\angstrom}$, enabling the determination of the P abundance. Additionally, it provides access to several other elements included in our analysis, namely C, N, O, Na, Mg, Al, Si, S, Ca, Fe, Ce, and Nd. For the abundance determination, we employed the Brussels Automatic Code for Characterizing High accUracy Spectra \cite[BACCHUS;][]{masseron2016} and relied on the stellar parameters provided by ASPCAP, the automated pipeline associated with APOGEE-2 \cite[][]{garciaperez16ASPCAP}. \\
The original group of 16 P-rich stars was discovered in the data of APOGEE-2 DR14. This group is included in the current study to update the results to the DR17 database. Furthermore, stars selected from a value added catalog (VAC) of the APOGEE-2 survey, which provided automated measurements of P in a large number of stars \cite[BAWLAS;][]{hayes22}, were added to the new sample for analysis. \\
By demonstrating that almost all Si-rich stars are also rich in P and verifying the P enhancements reported in the VAC, we expanded the sample to 78 P-rich giants, including the first detection of a P-rich star in the Galactic globular cluster M4. The left panel of Fig. \ref{P_Fe_P_Si} presents an update of Fig. 1 from \cite{masseron20a}, showing the [P/Fe] measurements of the expanded sample alongside literature values of [P/Fe] plotted against [Fe/H], which clearly demonstrate the enhancement. The differential approach chosen here—that is, the comparison of the abundances of P-rich stars with those of approximately 5\,500 'P-normal' stars with similar stellar parameters, analyzed using the same calculation procedure—also revealed several other overabundances, particularly in O, Al, Si, and Ce, as shown in Fig. \ref{abu_summary}. Systematic correlations among the enhanced elements, such as the proposed relationship between P and Si (see the right panel of Fig. \ref{P_Fe_P_Si}), were also identified. \\
Apart from the abundance analysis, we evaluated the possibility of binary mass transfer by inspecting the radial velocity scatter obtained from multiple observational visits of APOGEE-2. We also performed an orbital diagnostic with data from the astroNN VAC \cite[see][]{leung19VAC1,abdurrouf22APOGEEDR17}, which showed that the P-rich stars do not exclusively belong to any distinct local subgroup, such as accreted populations, indicating that they likely formed in situ as part of the Galactic thick disk and halo.

\section{Analyzing optical UVES spectra of four P-rich stars}\label{UVES}
To complement the chemical fingerprint obtained from the APOGEE-2 infrared data, we measured optical spectra of the four brightest P-rich stars using the Ultraviolet and Visual
Echelle Spectrograph \cite[UVES;][]{dekker2000} mounted at the Very Large Telescope (VLT). The spectra, which cover the wavelength range \unit[3300--9400]{\angstrom}, were analyzed using largely the same procedure as described in Sect. \ref{APOGEE}, employing the most recent line list compiled from the Vienna Atomic Line Database \cite[VALD;][]{piskonuv17, ryabchikova15}, which was carefully examined for excitation potential and log(gf). The broad wavelength coverage allowed the determination of the abundances of 48 light and heavy (Z $>$ 30) elements, ranging from Li to Pb, including rarely measured elements, such as those between Sr and Ag. Real measurements are accompanied by constraining upper limit estimates for Cd I, In I, and Sn I. \\
We found overabundances relative to solar in the s-process peak elements, along with an extremely high Ba abundance ([Ba/Fe] = \unit[0.88--1.84]{dex}) and slight enhancements in some elements between Rb and Sn. We also report measurements of the radial velocity and its scatter, obtained by combining observations from different nights, to reassess the possibility of a binary mass transfer scenario for the P-rich stars.

\begin{figure}[t]
	\centering
	\includegraphics[width=0.8\textwidth]{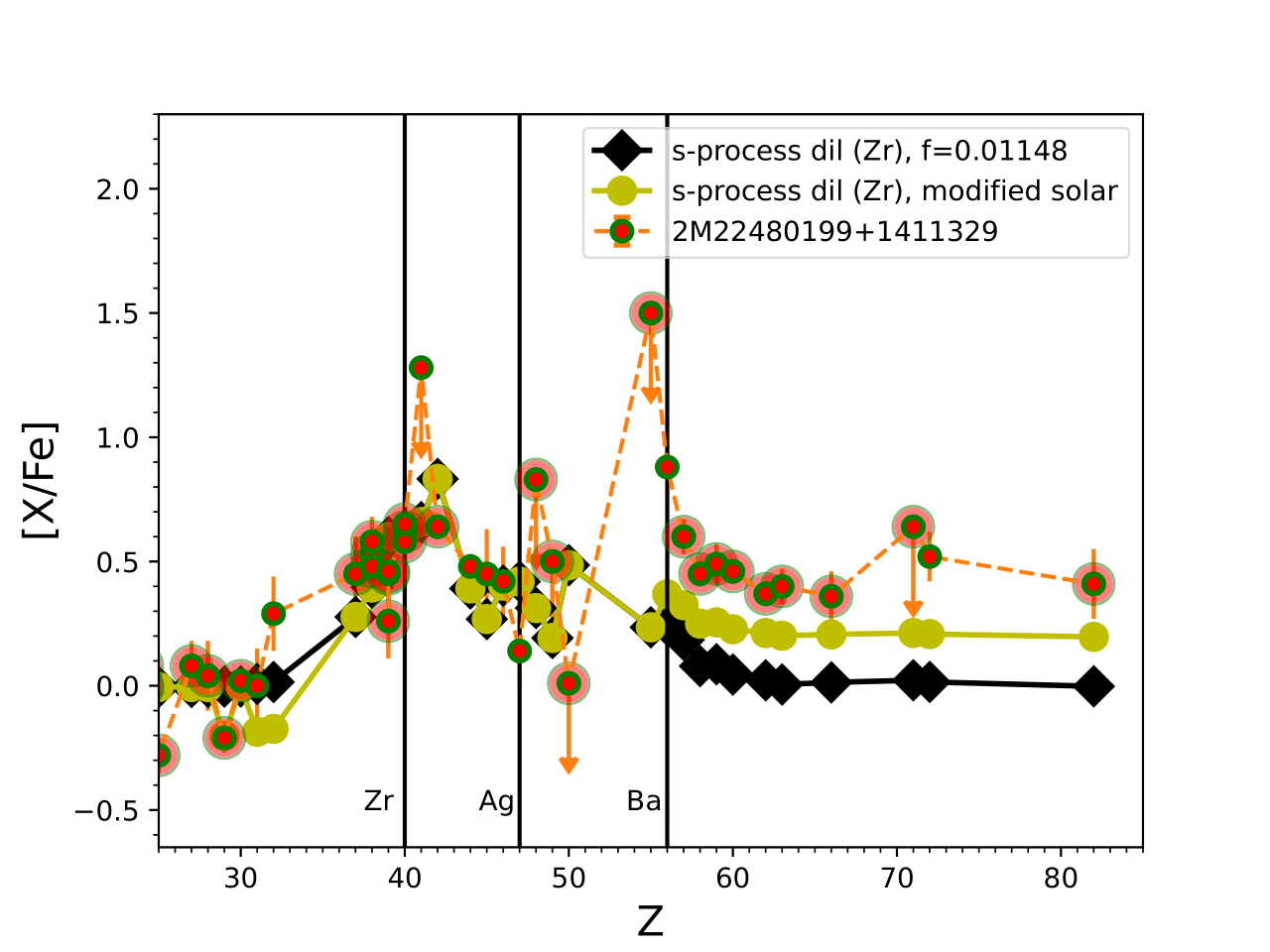} 
	\caption{Complete elemental abundances beyond Fe of 2M22480199+1411329 compared with s-process
	simulations from fast-rotating massive stars. The elements with the most
	reliable observational results (measurements or upper limits) are highlighted with large orange
	circles. Stellar simulations are diluted with pristine material to reproduce the abundance of
	the reference element Zr.}
	\label{s-process}
\end{figure}

\section{Searching for the progenitor of the P-rich stars}
Regarding binary interaction, we conclude from the radial velocity scatter of APOGEE-2 and UVES data that binarity could be a characteristic of some P-rich stars, but presumably, not all of them are part of close binary systems. It remains an open question for future long-term observations to definitively identify a binary P-rich star. Such an identification would shift the problem to the companion star, which transfers mass with a peculiar chemical pattern to the observed P-rich star; however, the origin of this pattern remains unknown. \\
Having defined the abundance signature of the P-rich stars (Fig. \ref{abu_summary}), we compared it with the typical pattern produced by specific nucleosynthetic events to find clues about the progenitor of the P-rich stars. Scenarios such as the thermonuclear explosion of carbon-oxygen white dwarfs (type Ia supernovae) with masses below the Chandrasekhar limit (subCh SNe Ia) and pair-instability supernovae (PISNe) can be ruled out due to mismatched abundances. In the case of subCh SNe Ia, the mismatch lies in the solar Ni abundance of the P-rich stars, instead of the subsolar value predicted by \cite{sanders21}, and in the absence of an odd-even effect, which should be prominent in the case of PISNe, as predicted by \cite{kozyreva14}. \cite{goriely22} proposed spallation processes by stellar energetic particles as an alternative to standard nucleosynthesis for the production of the abnormal abundances, but current spallation models underpredict the amounts of Al and Si, as well as the high [Ba/La] ratio. Furthermore, the exact conditions of such spallation events and the sites where they occur remain undefined. Very recently, \cite{bekki24} suggested that, under certain conditions, including only a small amount of mixing with the interstellar medium and the absence of strong pollution from core-collapse SNe (CCSNe) and SNe Ia products, newly formed stars could show a high [P/Fe] produced by oxygen-neon (ONe) novae. Previously, novae were tentatively ruled out by \cite{masseron20a} due to mismatched heavy-element signatures and the absence of C enhancement. The new models from \cite{bekki24} need to be tested against the more up-to-date results from Sect. \ref{UVES} to be properly evaluated. \\
In terms of nucleosynthetic processes, we know from previous works \cite[][]{masseron20a}, that the r-process, typical for CCSNe, is not compatible with the strange abundance pattern due to the relatively low Eu abundance. The same holds for the s-process that occurs in AGB stars, given the higher [Ba/La] ratio compared to that in AGB stars \cite[][]{masseron20b}. We performed a comparison with models of the nucleosynthetic s- and i-processes, for which the elements measured in Sect. \ref{UVES} play a crucial role. More precisely, we produced three scenarios using the post-processing network code PPN \cite[e.g.,][]{pignatari:12}: a single i- or s-process, a double i-process, and a combination of s- and i-processes. While some promising results have been found, none of the three models could provide a unique scenario reproducing the pattern of all four targets. The results shown in Fig. \ref{s-process}, which compares the abundance pattern of 2M22480199+1411329 with a single s-process from fast-rotating massive stars, confirm that the s-process cannot be the nucleosynthetic source of the P-rich stars. The specific stellar sites where the i-process may occur are not yet discovered, but possible hosts for the i-process include metal-poor evolved stars as well as massive stars, particularly during H-ingestion events in the He-burning layer. The currently limited capability of the 1D stellar models used to reproduce the behavior of those events introduces additional uncertainty, meaning that no definite conclusion can be drawn about the i-process from these models \cite[see, e.g.,][for a discussion]{clarkson:21}. \\ 
Ultimately, uncovering the progenitor of the P-rich stars remains an open task.
The discussion presented in this section, as well as the results from Sect. \ref{APOGEE} and \ref{UVES}, are available in \cite{brauner23} and \cite{brauner24}, respectively.

\section{Mining of new P-rich stars using Machine Learning}

\begin{figure}[t]
	\centering
	\includegraphics[width=0.49\textwidth]{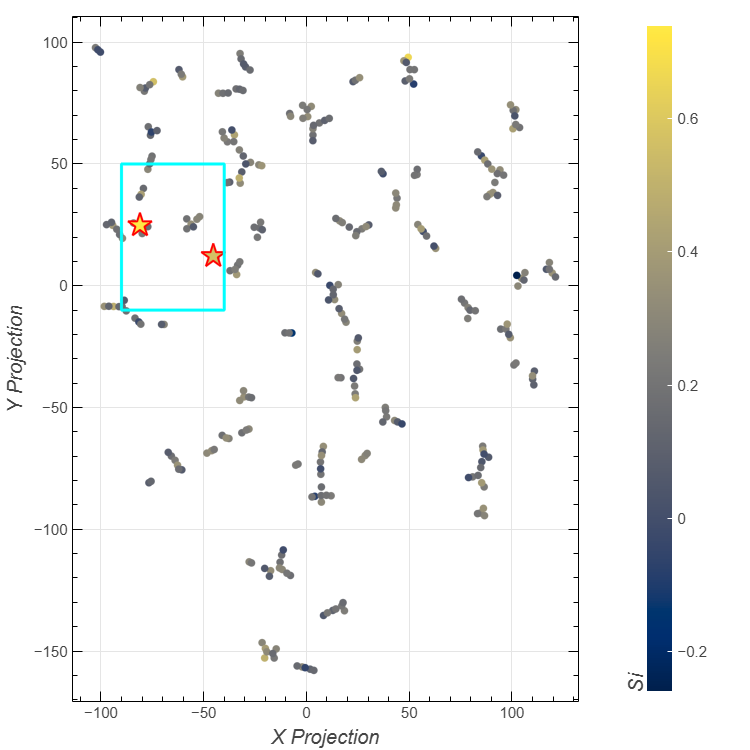} 
	\includegraphics[width=0.49\textwidth]{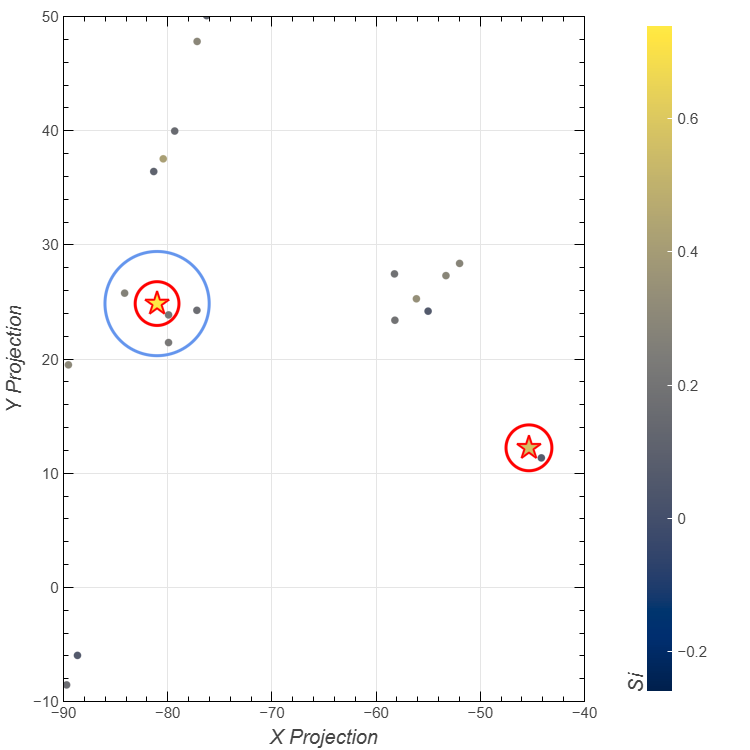}
	\caption{Output of the t-SNE run on the cut of the P line $\unit[15711.522]{\angstrom}$. Red star
		symbols highlight the P-rich stars. Left panel: Full output. Right panel: Zoom into the region selected in the left panel. Red circles include the stars selected as twins and blue circles include the second grade candidates.}
	\label{tSNE}
\end{figure}

As mentioned earlier, expanding the sample of known P-rich stars will help to define their chemical fingerprint with greater accuracy. The DR17 of APOGEE-2 provides spectra of over 650\,000 stars, and a systematic search through this extensive dataset may lead to the discovery of additional P-rich stars. Given the large volume of stellar spectra, we opted for unsupervised machine learning approaches for this task, starting with a popular clustering algorithm called t-distributed stochastic neighbor embedding \cite[t-SNE;][]{t-SNE}. In short, t-SNE takes high-dimensional feature vectors as input and reduces their dimensionality. This reduction facilitates visualization by embedding the data into a two-dimensional space, preserving similarities between vectors in the low-dimensional space and resulting in clustering. To discover P-rich stars in the APOGEE-2 spectra, we define the feature vector as the flux values of the P line and/or Si lines. Since stellar parameters such as [M/H] and T$_{\textit{eff}}$ strongly influence the morphology of the spectra, we first group the stars into bins based on [M/H] and T$_{\textit{eff}}$ before applying the t-SNE clustering. This step maximizes the abundance information used when grouping stars by similarity, helping to identify spectral twins of the P-rich stars. An example of this identification is shown in Fig. \ref{tSNE}, which displays the output of a t-SNE run for a single parameter bin. Stars located near the P-rich star in the t-SNE visualization are extracted and categorized as either twins or second-grade candidates, depending on their proximity. This approach introduces a priority ranking for the stars to be analyzed in detail, following the methodology used in Sect. \ref{APOGEE}. \\
Other techniques, such as the use of Hierarchical Agglomerative Clustering (HAC) and the application of t-SNE on the output of a disentangling auto-encoder \cite[see][]{Santovena24}, designed to remove the influence of stellar parameters on the spectra through the use of neural networks, eliminating the need for prior binning, have also been explored. All three approaches have identified a large number of P-rich candidates. For instance, the application of t-SNE on the strongest P I line $\unit[15711.522]{\angstrom}$, combined with prior binning, yielded 72 twins and 148 second-grade candidates. However, final confirmation depends on future detailed chemical abundance analyses, such as those conducted in Sect. \ref{APOGEE}. \\ \\


MB acknowledges support from the Fundación "la Caixa" and the grant PRE-2020-095531 of the Severo Ochoa Program for the Training of Pre-Doc Researchers (FPI-SO) from the European Union and the State Agency of Investigation of the Spanish Ministry of Science and Innovation  (MICINN). \\
MP thanks the support from the NKFI via K-project 138031 and the Lend\"ulet Program LP2023-10 of the Hungarian Academy of Sciences. MP acknowledge the support to NuGrid from IReNA, supported by US NSF AccelNet (Grant No. OISE-1927130) and from the European Union’s Horizon 2020 research and innovation programme ChETEC-INFRA (Project no. 101008324).


\end{document}